\documentclass[preprint,12pt]{elsarticle}

\usepackage[utf8]{inputenc}
\usepackage{graphicx}
\usepackage{caption}           
\usepackage{subcaption}        
\usepackage[inline]{enumitem}  
\usepackage{csquotes}
\usepackage{epstopdf}
\usepackage{setspace}
\usepackage{ragged2e}
\usepackage{amsmath}
\usepackage{amsthm}
\usepackage{amssymb}
\usepackage{cleveref}

\usepackage{graphicx, color}
\graphicspath{{fig/}}
\usepackage{algorithm, algpseudocode} 
\usepackage{mathrsfs} 
\def\R{{\mathbb R}}

\usepackage{geometry} 
\usepackage{multirow}
\usepackage[table,xcdraw]{xcolor}

\newcommand\zqp [1]{\textcolor{black}{#1}}   
\newcommand\tclr [1]{\textcolor{black}{#1}}   

\usepackage{lineno} 
\journal{Expert Systems}
\begin{document}

\begin{frontmatter}
\title{MCDIP-ADMM: Overcoming Overfitting in DIP-based CT reconstruction}
\author[label1]{Chen Cheng}
\ead{chengchen0301@sjtu.edu.cn}
\author[label2]{Qingping Zhou\corref{cor1}}
\cortext[cor1]{Corresponding author}
\ead{qpzhou@csu.edu.cn}
\affiliation[label1]{organization={School of Mathematical Sciences, Shanghai Jiao Tong University},
            city={Shanghai},
            postcode={200240}, 
            country={China}}
\affiliation[label2]{organization={School of Mathematics and Statistics, HNP-LAMA, Central South University},
            city={Changsha},
            postcode={Hunan 410083}, 
            country={China}}
\begin{abstract}
This paper investigates the application of unsupervised learning methods for computed tomography (CT) reconstruction.
To motivate our work, we review several existing priors, \zqp{namely the truncated Gaussian prior, the $l_1$ prior, the total variation prior, and the deep image prior (DIP).} We find that DIP \zqp{outperforms the other three priors in terms of representational capability and visual performance.
However, the performance of DIP deteriorates when the number of iterations exceeds a certain threshold due to overfitting.} To address this issue, \zqp{we propose a novel method (MCDIP-ADMM) based on Multi-Code Deep Image Prior and plug-and-play Alternative Direction Method of Multipliers.
Specifically, MCDIP utilizes multiple latent codes to generate a series of feature maps at an intermediate layer within a generator model. These maps are then composed with trainable weights, representing the complete image prior.} \zqp{Experimental results demonstrate the superior performance of the proposed MCDIP-ADMM compared to three existing competitors. In the case of parallel beam projection with Gaussian noise, MCDIP-ADMM achieves an average improvement of 4.3 dB over DIP, 1.7 dB over ADMM DIP-WTV, and 1.2 dB over PnP-DIP in terms of PSNR. Similarly, for fan-beam projection with Poisson noise, MCDIP-ADMM achieves an average improvement of 3.09 dB over DIP, 1.86 dB over ADMM DIP-WTV, and 0.84 dB over PnP-DIP in terms of PSNR.} 
\\
\end{abstract}

\end{frontmatter}

\newpage
\setcounter{page}{1}
\section{Introduction}
Computed tomography (CT) is a widely used medical imaging modality, with various applications in clinical settings, such as diagnostics, screening, and virtual treatment planning, as well as in industrial and scientific settings~\cite{kofler2018u,fan2022model,leuschner2021lodopab}.
In the CT restoration, it would be required to \zqp{keep the number of angles} as low as possible because excessive doses of applied radiation might be detrimental to patients. \zqp{The sparse measurements pose a significant challenge to obtaining high-quality reconstruction.}
Furthermore, a common challenge in these problems is their ill-posedness due to measurement noise and undersampling, which indicates that multiple consistent images map to the same observations. As a result, \zqp{prior knowledge of the image is essential to determine the most probable solutions among these potential images}.

\zqp{
In recent years, deep learning has been successfully applied to CT reconstruction problems, often through a supervised learning scheme \cite{chen2017low, adler2018learned,he2018optimizing}. Convolutional neural networks (CNNs) are commonly used to learn data-driven priors in these approaches \cite{guo2023physics}. However, supervised learning typically requires a large amount of paired training data to be effective. Deep image prior (DIP) provides an unsupervised alternative for image reconstruction using deep learning \cite{ulyanov2018deep,nittscher2023svd}. One of its main advantages over supervised methods is that it does not require any training data and learns solely from the observed data sample, relying on the inherent structure of CNNs to impose regularization on the image. In general, DIP approaches are often overparameterized, leading to overfitting and a decline in quality when the iteration exceeds a certain point \cite{shi2022measuring,ding2021rank,wang2021early,you2020robust}. Plug-and-play DIP (PnP-DIP) \cite{sun2021plug} is a variant of DIP that addresses the issue of overfitting. It incorporates a plug-and-play prior scheme that allows for additional regularization steps within the DIP framework. However, PnP-DIP still encounters challenges with overfitting when handling highly ill-posed inverse problems such as denoising.
}

\zqp{
The problem of overfitting is not unique to DIP-based reconstruction. Recently, Gu et al.~\cite{gu2020image} proposed a novel way of adapting 
multiple latent codes of a generative adversarial network (GAN) as an effective prior to mitigate overfitting. They demonstrated the efficacy of this approach in various image processing tasks such as image colorization, super-resolution, image inpainting, and semantic manipulation.  
In this paper, our contribution is to integrate the idea of the multiple latent codes into the DIP framework, which leads to remarkable stability.
Specifically, we introduce a multi-code deep image prior (MCDIP) based on a generative model. We then incorporate MCDIP into the plug-and-play ADMM optimization. 
Through our experiments, we have observed that the MCDIP-ADMM approach demonstrates remarkable resistance to overfitting in the presence of both parallel beam projection with Gaussian noise and fan-beam projection with Poisson noise. These findings highlight the robustness of MCDIP-ADMM in handling challenging scenarios and further validate its effectiveness in addressing the overfitting issue in CT reconstruction.
}

The rest of the paper is organized as follows. Section~\ref{sec:motivation} introduces the imaging inverse problem before reviewing and visualizing some of the most commonly used image priors for comparison. Section~\ref{sec:dip-pnp} introduces the plug-and-play deep image prior, while Section~\ref{sec:propose} presents the proposed method. Section~\ref{sec:experiments} presents numerical experiments. Section~\ref{sec:conclusion} concludes with a discussion about the future directions of this research.

\section{Background and motivation}\label{sec:motivation}
\paragraph{\textbf{Image inverse problems}}  We start by presenting the formulation of the generic image inverse problem. Let $X,Y$ be Hilbert or Banach spaces with inner product $\langle\cdot, \cdot\rangle$, data $\boldsymbol{y} \in Y \subset \R^{m}$ is related to $\boldsymbol{x} \in X \subset \R^n$ via the following imaging process:
\[ \boldsymbol{y} =A \boldsymbol{x} \oplus \tau, \]
where $A$ is the measurement operator that models the image acquisition process. It corresponds to an identity operator in the denoising problem and a sub-sampled radon transform in the CT reconstruction problem, to name a few. $\tau$ describes the noise in the measurement, with $\|\tau\| \leqslant \delta$. When Gaussian noise is assumed, the operator $\oplus$ implies addition; when Poisson noise is assumed, it denotes a nonlinear operator. The aim is to obtain an approximation $\hat{\boldsymbol{x}}$ for the true solution $\boldsymbol{x}$.

Classical methods of solving this inverse problem involve a regularized optimization given by:
\begin{equation}\label{e:optim_orig}
 \hat{\boldsymbol{x}}  = \arg \min_{\boldsymbol{x}} \mathcal{F}\left(A \boldsymbol{x}, \boldsymbol{y} \right)+ \mathcal{R} (\boldsymbol{x}).
\end{equation}
The term $\mathcal{F}\left(A \boldsymbol{x}, \boldsymbol{y} \right)$ is fidelity loss that measures the consistency of the approximate solution to the observed data $\boldsymbol{y}$.
The second component $\mathcal{R} (\boldsymbol{x})$ is a regularization term that represents one's prior knowledge of the unknown image. 

\paragraph{\textbf{Image prior}}  
\zqp{
The selection of an appropriate regularizer is a crucial step in modeling and has garnered considerable attention in the literature~\cite{kaipio2006statistical,ulyanov2018deep,mataev2019deepred}.
Traditional optimization methods rely on handcrafted priors modeled as convex regularization functions such as the total variation (TV), encouraging smoothness, or the $l_1$ norm for sparsity~\cite{kaipio2006statistical}. In this context, convex optimization algorithms have played an important role and their convergence properties have been well established. 
However, the hand-crafted priors are now significantly outperformed by deep neural networks-based priors, which directly learn the inverse mapping from degraded measurements to the solution space. 
One such powerful prior is the deep image prior (DIP), which relies on using untrained networks such as a generator or a U-net and optimizing their weights directly during the reconstruction process.} We review the aforementioned classical priors in the application of image reconstruction methods and obtain intuitive results. 
The density of Gaussian white noise with positivity constraint is
\begin{equation}\label{e:gaussian_prior}
 \pi_{\mathrm{TG}}(\boldsymbol{x}) \propto \pi_{+}(\boldsymbol{x}) \exp \left(-\frac{1}{2 \alpha^{2}}\|\boldsymbol{x}\|^{2}\right), 
\end{equation}
where $\pi_{+}(\boldsymbol{x})=1$ if $x_{j}>0$ for all $j$  and $\pi_{+}(\boldsymbol{x})=0$ otherwise. The $\|\cdot\|$ is the usual Euclidian norm. The $\ell^{1}$ prior with positivity constraint is defined as
\begin{equation}\label{e:l1_prior}
\pi_{\ell^{1}}(\boldsymbol{x})=\alpha^{n} \pi_{+}(\boldsymbol{x}) \exp \left(-\alpha\|\boldsymbol{x}\|_{1}\right),
\end{equation}
where $\|\boldsymbol{x}\|_{1} = \sum_{i,j}^{n}\left|x_{ij}\right|$ and $\alpha$ is a hyperparameter. The total variation of the discrete image $\boldsymbol{x}=\left[x_{1}, x_{2}, \ldots, x_{n}\right]^{\mathrm{T}}$ is now defined as
\begin{equation}\label{e:tv_def}
\mathrm{TV}(\boldsymbol{x})=\sum_{j=1}^{n} V_{j}(\boldsymbol{x}), \quad V_{j}(\boldsymbol{x})=\frac{1}{2} \sum_{i \in \mathcal{N}_{j}} \ell_{i j}\left|x_{i}-x_{j}\right|, 
\end{equation}
where $\mathcal{N}_{i}$ is the index set of the neighbors of $x_i$ and $\ell_{i j}$ is the length of the edge between the neighboring pixels. The discrete total variation density will now be given as
\begin{equation}\label{e:tv_prior}
\pi_{\mathrm{TV}}(\boldsymbol{x}) \propto \exp (-\alpha \operatorname{TV}(\boldsymbol{x})),
\end{equation}
where $\alpha$ is a positive constant.
\begin{figure}[!htb]
\centering
  
  \begin{subfigure}[t]{0.24\textwidth}
    \centering
    \includegraphics[width=0.99\linewidth]{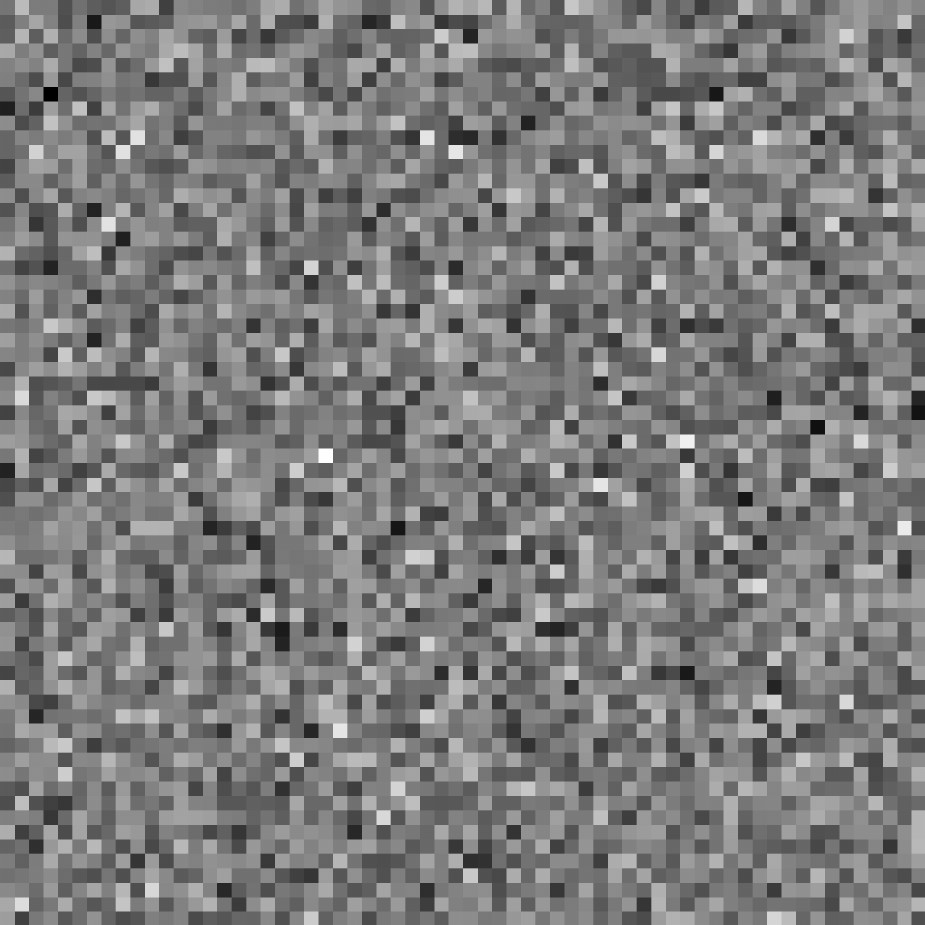}
    \caption{$x_{\mathrm{TG}}$}
  \end{subfigure}
  \begin{subfigure}[t]{0.24\textwidth}
    \centering
    \includegraphics[width=0.99\linewidth]{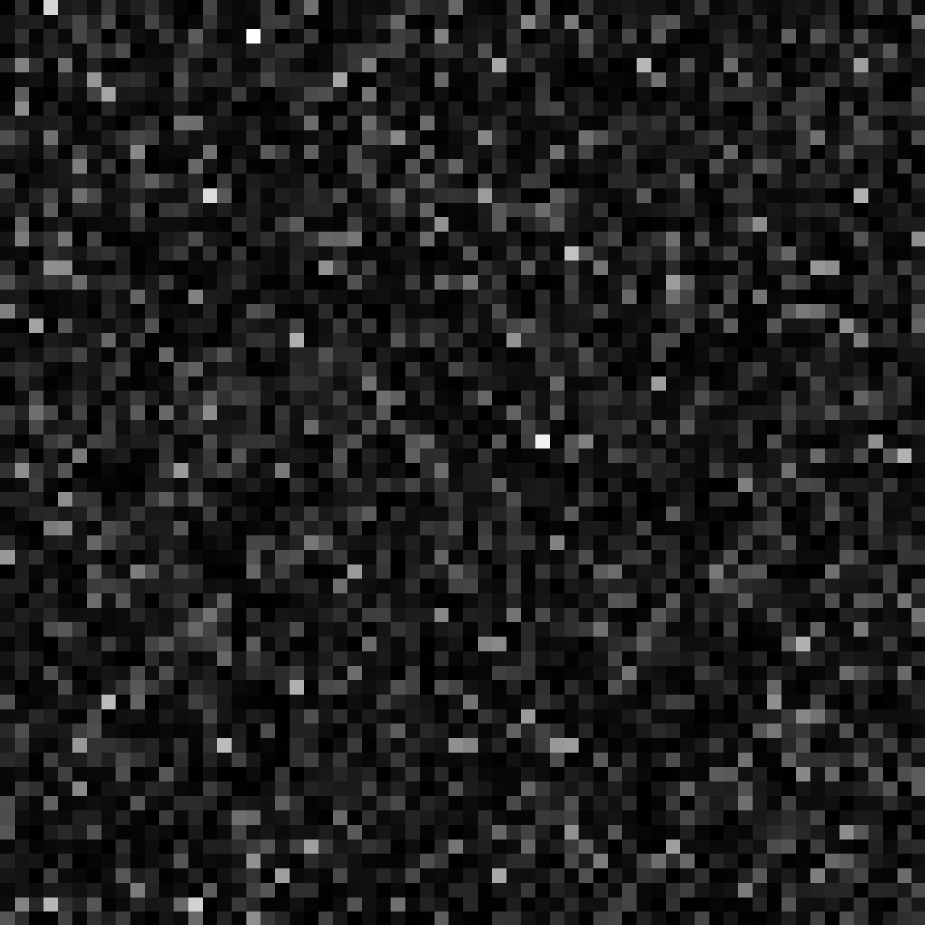}
    \caption{$x_{l_1}$}
  \end{subfigure}
  \begin{subfigure}[t]{0.24\textwidth}
    \centering
    \includegraphics[width=0.99\linewidth]{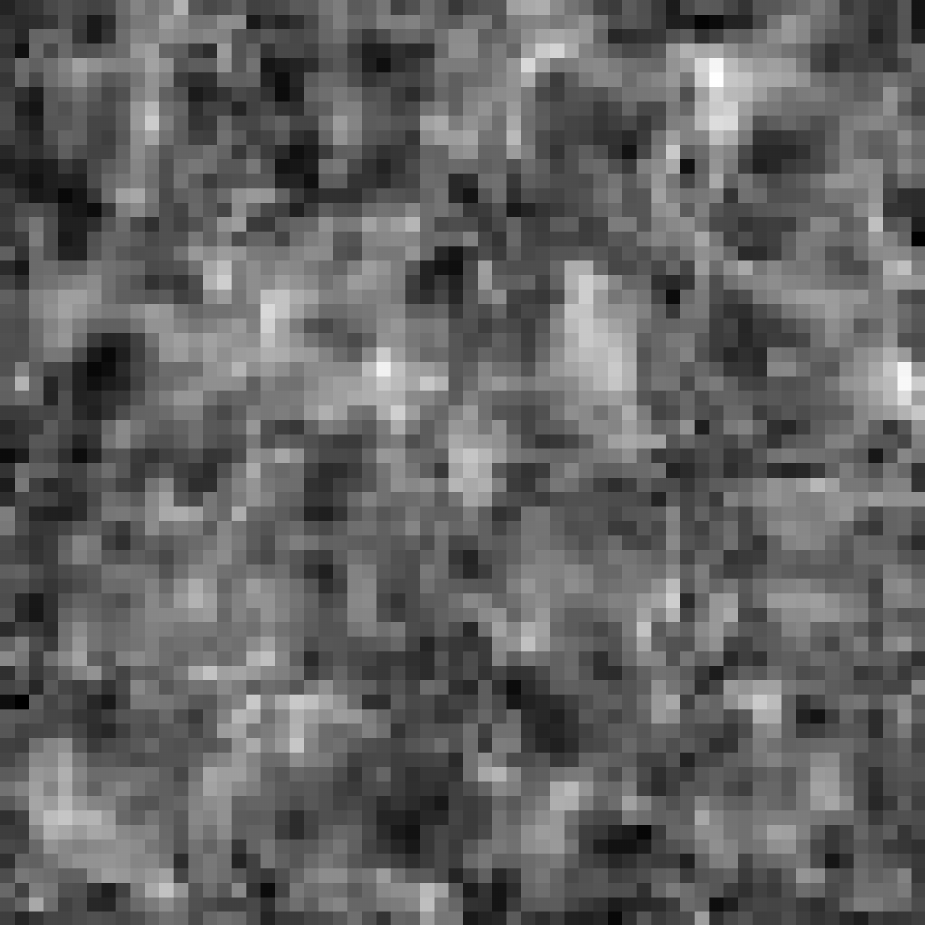}
    \caption{$x_{\mathrm{TV}}$}
  \end{subfigure}
    \begin{subfigure}[t]{0.24\textwidth}
    \centering
    \includegraphics[width=0.99\linewidth]{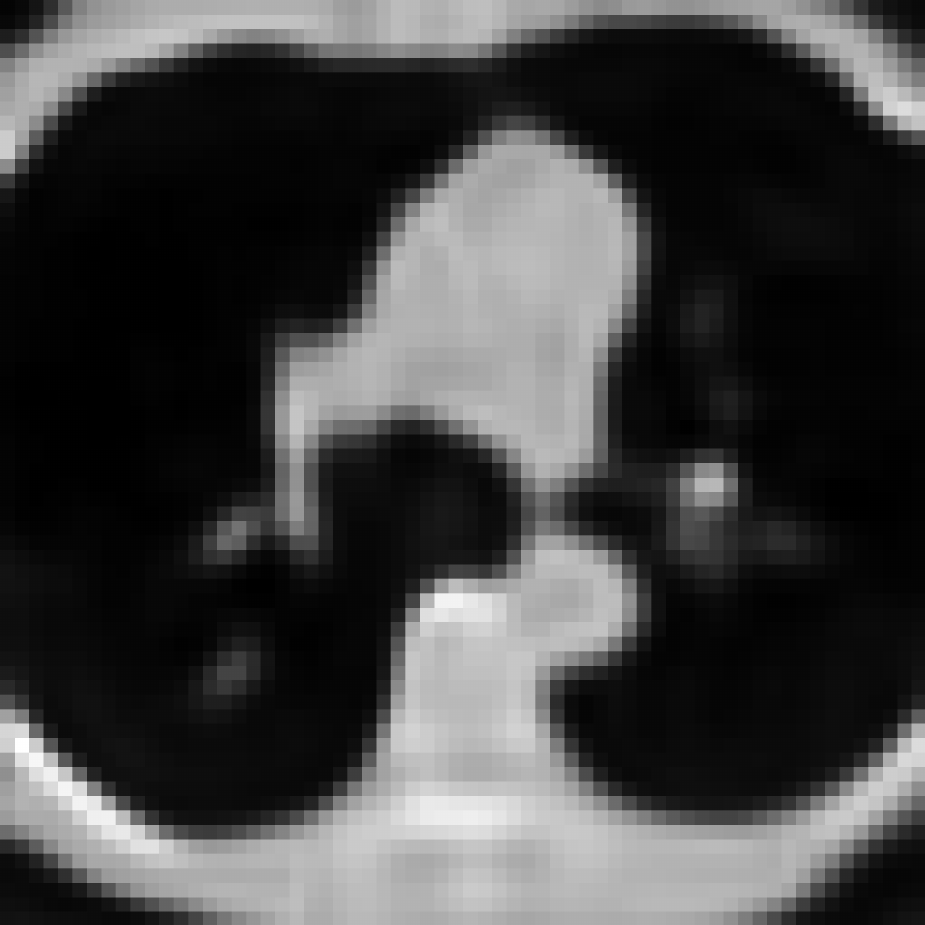}
    \caption{$x_{\mathrm{DIP}}$}
  \end{subfigure}
  \caption{\tclr{One random sample from Gaussian white noise prior in Eq.~\eqref{e:gaussian_prior}, $\ell^1$ prior in Eq.~\eqref{e:l1_prior}, total variation prior in Eq.~\eqref{e:tv_prior} and deep image prior, respectively. From left to right, $x_{TG} \sim \pi_{\mathrm{TG}}(\cdot)$, $x_{l1} \sim \pi_{\ell^{1}}(\cdot)$, $x_{tv} \sim \pi_{\mathrm{TV}}(\cdot)$, and $x_{\mathrm{DIP}}$ is the output of a randomly initialized DIP model with the observed measurement as input.} }
  \label{f:samples_prior}
\end{figure}

In Fig.~\ref{f:samples_prior}, we describe $64 \times 64$ noise images randomly drawn from distributions of the Gaussian prior, $\ell^{1}$ prior, and the TV prior. The parameter $\alpha$ in the priors is chosen all equal to unity. For comparison, a sample from deep image prior (DIP)~\cite{ulyanov2018deep} is also displayed. As seen in Fig.~\ref{f:samples_prior}, $\ell^1$ prior can depict discontinuities, i.e., for some big leaps in the image. Besides sharp leaps, TV prior also exhibits a clear correlation between nearby pixels. With its edge-preserving property, it has become the most commonly used prior in the image reconstruction field. A deep image prior sample, on the other hand, contains both the qualitative and quantitative features of the image, distinguishing it from other priors.

\section{Plug-and-Play deep image prior} \label{sec:dip-pnp}
\zqp{We start with a concise overview of the plug-and-play deep image prior (PnP-DIP) following the presentation of~\cite{sun2021plug}.}
The reconstructed image is parameterized as $\boldsymbol{x} = G(\boldsymbol{z};\theta)$, where $G$ represents a deep generative network with randomly initialized weights $\theta \in \Theta$ and a fixed input $\boldsymbol{z}$ such as random white noise. The optimization formation can be written as follows: 

\begin{equation}\label{e:pnp_ori}
\begin{aligned}
\min\limits_{\boldsymbol{x}, \theta}& \quad \mathcal{F}\left(A \boldsymbol{x}, \boldsymbol{y} \right) \\
\text{s.t.}& \quad \boldsymbol{x}=G(\boldsymbol{z}; \theta)
\end{aligned}   
\end{equation}
The problem can then be solved by optimizing over the weights of the generator $G(\boldsymbol{z}; \theta)$ by a gradient descent method to minimize the data discrepancy of the network's output.
The overfitting issue observed when using DIP suggests that the regularization term $\mathcal{R} (\boldsymbol{x})$ is required. On the other hand, DIP may incorporate state-of-the-art regularized  methods as priors in more challenging inverse problems. Because these regularizers are non-differentiable, they can not be solved using the vanilla gradient descent method in the form of an optimization problem; instead, the problem is reformulated by splitting the optimization variables into $\boldsymbol{x}$ and $\boldsymbol{w}$: 
\begin{equation}
\begin{aligned}\label{e:PnP}
\min\limits_{\boldsymbol{w}, \boldsymbol{x}}& \quad \mathcal{F}\left(A \boldsymbol{w}, \boldsymbol{y} \right)+\mathcal{R}(\boldsymbol{x}) \\
\text{s.t.}& \quad \boldsymbol{x}=\boldsymbol{w}
\end{aligned}
\end{equation}
Since both two splitted variables are in two separate terms, the reformulated problem can be solved by altering direction method of multipliers(ADMM). Then by replacing the variable $\boldsymbol{w}$ with a network $G(\boldsymbol{z}; \theta)$, the problem is reformulated in a hybrid version of DIP and plug-and-play priors~\cite{venkatakrishnan2013plug}:
$$
\begin{aligned}
\min\limits_{\theta, \boldsymbol{x}}& \quad \mathcal{F}\left(A G(\boldsymbol{z};\theta), \boldsymbol{y} \right)+\mathcal{R}(\boldsymbol{x}) \\
\text{s.t.}& \quad \boldsymbol{x}=G(\boldsymbol{z} ; \theta)
\end{aligned}
$$
The augmented Lagrangian of this new problem can be written as 
$$
\begin{aligned}
\mathcal{L}_{\rho}(\boldsymbol{x}, \theta, \boldsymbol{u})=& \mathcal{F}\left(A G(\boldsymbol{z};\theta), \boldsymbol{y} \right)+\mathcal{R}(\boldsymbol{x}) \\ &+\frac{\rho}{2}\|\boldsymbol{x}-G(\boldsymbol{z} ; \theta)+\boldsymbol{u}\|_{2}^{2}-\frac{\rho}{2}\|\boldsymbol{u}\|_{2}^{2},
\end{aligned}
$$
where $\boldsymbol{u}$ is the dual variable and $\rho$ is the penalty parameter. The corresponding ADMM steps for this augmented Lagrangian are
$$
\begin{aligned}
\boldsymbol{x}_{t+1}&={\operatorname{prox}_{\frac{R}{\rho}}}\left(G\left(\boldsymbol{z} ; \theta_{t}\right)-\boldsymbol{u}_{t}\right) \\
\theta_{t+1}&=\underset{\theta}{\operatorname{argmin}} \: \mathcal{L}_{\rho}\left(\boldsymbol{x}_{t+1}, \theta, \boldsymbol{u}_{t}\right) \\
\boldsymbol{u}_{t+1}&=\boldsymbol{u}_{t}+\left(\boldsymbol{x}_{t+1}-G\left(\boldsymbol{z} ; \theta_{t+1}\right)\right),
\end{aligned}
$$
which correspond to an $\boldsymbol{x}$-minimization step, a $\theta$-minimization step and a dual ascent step respectively.

\section{Overcoming Overfitting via MCDIP} \label{sec:propose}
\subsection{MCDIP}
\begin{figure}[t]
\raggedleft
\includegraphics[width=1.0\textwidth]{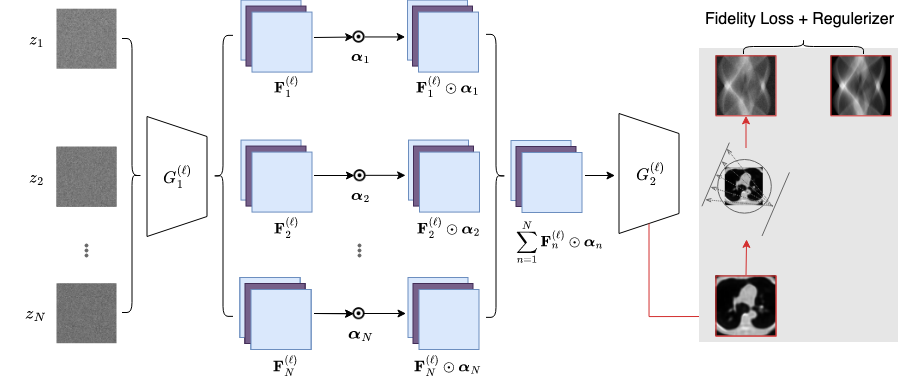}
\caption{\tclr{Pipeline of the proposed MCDIP-ADMM approach. The network $G(\cdot)$  of DIP is divided into two sub-networks $G_{1}^{(\ell)}(\cdot)$ and $G_{2}^{(\ell)}(\cdot)$. By employing multiple latent codes $\left\{\boldsymbol{z}_{n}\right\}_{n=1}^{N}$, we then obtain the restoration $\hat{\boldsymbol x} = G_{2}^{(\ell)}\left(\sum_{n=1}^{N} G_{1}^{(\ell)}\left(\boldsymbol{z}_{n}\right) \odot \boldsymbol{\alpha}_{n}\right)$.} }
\label{f:multi-mode}
\end{figure}

\zqp{
Considering the limited representational capacity of networks with a single latent code, we instead employ a generative network with multiple inputs, serving as the deep image prior, as outlined in Fig.\ref{f:multi-mode}. The multi-code deep image prior (MCDIP) utilizes multiple latent codes to generate several feature maps at an intermediate layer within the generative network, which is then composed using learned weights.
We partition the generative network $G(\cdot)$ into two sub-networks: $G_{1}^{(\ell)}(\cdot)$ and $G_{2}^{(\ell)}(\cdot)$.} Here, $\ell$ is the index of the immediate layer to perform feature composition. For each $\boldsymbol{z}_n$, such a separation can extract the corresponding spatial feature 
\[\mathbf{F}_n^{(\ell)} = G_{1}^{(\ell)}(\boldsymbol{z}_n)\] 
for further composition. 
Adaptive channel importance $\boldsymbol{\alpha}_{n} \in \mathbb{R}^{C}$ is introduced for each $\boldsymbol{z}_n$ to help them align with different semantics. Here, $C$ is the number of channels in the $\ell$-th layer of $G(\cdot)$ and $\boldsymbol{\alpha}_{n}$ is a $C$-dimensional vector representing the importance of the corresponding channel of the feature map $\mathbf{F}_n^{(\ell)}$. With such a composition, the image can be generated with
\begin{equation}\label{e:multi-model}
G(\boldsymbol{z}_1, ..., \boldsymbol{z}_n; \theta)=G_{2}^{(\ell)}\left(\sum_{n=1}^{N} \mathbf{F}_n^{(\ell)} \odot \boldsymbol{\alpha}_{n}\right),
\end{equation}
where $\odot$ denotes the channel-wise multiplication as
\begin{equation}
\left\{\mathbf{F}_n^{(\ell)} \odot \boldsymbol{\alpha}_{n}\right\}_{i, j, c}=\left\{\mathbf{F}_n^{(\ell)}\right\}_{i, j, c} \times\left\{\boldsymbol{\alpha}_{n}\right\}_{c}.
\end{equation}
Here, $i$ and $j$ indicate the spatial location, while $c$ stands for the channel index.

\subsection{MCDIP-ADMM}
Substituting~\eqref{e:multi-model} into~\eqref{e:optim_orig}, the optimization problem can therefore be written as follows:
\begin{equation}\label{e:prop_obj}
\begin{aligned}
\min\limits_{\theta, \boldsymbol{x}}& \quad \mathcal{F}\left(A G(\boldsymbol{z}_1, ..., \boldsymbol{z}_n;\theta), \boldsymbol{y} \right)+\mathcal{R}(\boldsymbol{x}) \\
\text{s.t.}& \quad \boldsymbol{x}=G(\boldsymbol{z}_1, ..., \boldsymbol{z}_n; \theta)
\end{aligned}
\end{equation}
where the network $G(\boldsymbol{z}_1, ..., \boldsymbol{z}_n, \theta)$ with multiple input features is defined as in Eq.~(\ref{e:multi-model}), and $\mathcal{R}(\boldsymbol{x})=\lambda \mathrm{TV}(\boldsymbol{x})$, where $\mathrm{TV}(\cdot)$ is defined in Eq.~(\ref{e:tv_def}). The regularization parameter $\lambda$ determines the amount of the contribution of each term. If $\lambda$ is large enough, the regularization term contributes more to the minimization issue than the data fidelity term. On the other hand, if $\lambda$ is small enough, it frequently means that the regularizer is so weak that the data dominates the minimization issue. This circumstance is particularly undesirable in the ill-posed medical image reconstruction issue, as the problem is highly ill-posed and we require a large contribution from the regularization term to produce accurate estimates of the unknown picture. In this regard, we should pick $\lambda$ in such a manner that the impacts of the regularization term and the data term are properly balanced.

The new formation is also an updated version of DIP prior that the generator $G(\boldsymbol{z};\theta)$ in DIP is substituted with the multi-code version of  network $G(\boldsymbol{z}_1, ..., \boldsymbol{z}_n; \theta)$. As a result, it retains the benefits of PnP-DIP in terms of limiting the overfitting issue observed when using DIP, and is hopeful of outperforming PnP-DIP thanks to the improved expressiveness of the network $G(\boldsymbol{z}_1, ..., \boldsymbol{z}_n; \theta)$. To solve this new problem, we first write the augmented Lagrangian with scaled dual variable $u$ and penalty parameter $\rho$ in the following form:
$$
\begin{aligned}
\mathcal{L}_{\rho}(\boldsymbol{x}, \theta, \boldsymbol{u})=& \mathcal{F}\left(A G(\boldsymbol{z}_1, ..., \boldsymbol{z}_n;\theta), \boldsymbol{y} \right)+\mathcal{R}(\boldsymbol{x}) \\
&+\frac{\rho}{2}\|\boldsymbol{x}-G(\boldsymbol{z}_1, ..., \boldsymbol{z}_n; \theta)+\boldsymbol{u}\|_{2}^{2}-\frac{\rho}{2}\|\boldsymbol{u}\|_{2}^{2},
\end{aligned}
$$
the optimization problem can be solved by the following ADMM steps:
$$
\begin{aligned}
\boldsymbol{x}_{t+1}&={\operatorname{prox}_{\frac{R}{\rho}}}\left(G\left(\boldsymbol{z}_1, ..., \boldsymbol{z}_n; \theta_{t}\right)-\boldsymbol{u}_{t}\right) \\
\theta_{t+1}&=\underset{\theta}{\operatorname{argmin}} \: \mathcal{L}_{\rho}\left(\boldsymbol{x}_{t+1}, \theta, \boldsymbol{u}_{t}\right) \\
\boldsymbol{u}_{t+1}&=\boldsymbol{u}_{t}+\left(\boldsymbol{x}_{t+1}-G\left(\boldsymbol{z}_1, ..., \boldsymbol{z}_n; \theta_{t+1}\right)\right).
\end{aligned}
$$
It should be noted that the $\theta$-minimization step is often intractable. For computational efficiency, we take one gradient descent step for $\theta$. This simple first-order update is sufficient to obtain good results in our experiments. Our method is summarized in Algorithm~\ref{alg:1}. \zqp{
While classical plug-and-play frameworks~\cite{venkatakrishnan2013plug,chan2016plug, lai2022deep}
aim to replace the proximal regularization term with an advanced denoising algorithm, our focus lies in utilizing  MCDIP as a plug-and-play prior, while simultaneously retaining the proximal regularization term.
}

\zqp{
We abstain from using early stopping for MCDIP and its comparisons due to two main reasons. First, finding simple 
 strategies for early stopping is challenging, as  evidenced by the extensive literature  on this topic. Most early-stopping approaches are tailored to specific models or problems~\cite{shi2022measuring,wang2021early}, and frequently fail to generalize across different settings. Second, our main focus is to showcase a method with exceptional stability that circumvents the necessity for dependable early stopping criteria.}

\begin{algorithm}[t]
\caption{MCDIP-ADMM} 
\label{alg:1}
{\bf Input:} 
Augmented Lagrangian $\mathcal{L}_{\rho}$, Regularizer $R$, Convolutional Neural Network $G$, Number of latent codes $N$, Number of iterations $T$, Step size for $\theta$: $\beta$, Optimizer for $\theta$: Opt
\smallskip
\begin{algorithmic}[1]
\State Sample $N$ latent codes $\left\{\boldsymbol{z}_{n}\right\}_{n=1}^{N}$ from Gaussian distribution
\State Initialize $\theta_0$, $\boldsymbol{x}_0$, $\boldsymbol{u}_0$
\For{$t = 0, 1, ..., T-1$}
    \State $\boldsymbol{x}_{t+1} = \operatorname{prox}_{\frac{R}{\rho}}\left(G\left(\boldsymbol{z}_1, ..., \boldsymbol{z}_n; \theta_{t}\right) - \boldsymbol{u}_t\right) \leftarrow$ 
  proximal step for $\boldsymbol{x}$
    \State $g_{\theta} = \nabla_{\theta} \mathcal{L}_{\rho}\left(\boldsymbol{x}_{t+1}, \theta, \boldsymbol{u}_{t}\right) \leftarrow$ calculate gradient w.r.t. $\theta$
    \State $\theta_{t+1} = \theta_{t}-\beta \cdot \operatorname{Opt}\left(\theta_{t}, g_{\theta}\right) \leftarrow$ update $\theta$ using gradient-based optimizer
    \State $\boldsymbol{u}_{t+1}=\boldsymbol{u}_{t}+\left(\boldsymbol{x}_{t+1}-G\left(\boldsymbol{z}_1, ..., \boldsymbol{z}_n; \theta_{t+1}\right)\right) \leftarrow$ dual ascent step 
\EndFor
\end{algorithmic}
{\bf Output:} reconstruction image $\boldsymbol{x}_T$
\end{algorithm}

\section{Experiments and discussion} \label{sec:experiments}
\zqp{In this section, we present the experimental results of low-dose CT reconstruction using parallel beam and fan-beam projections. We compare our method against three approaches: DIP~\cite{ulyanov2018deep}, ADMM DIP-WTV~\cite{cascarano2021combining}, and PnP-DIP~\cite{sun2021plug}.}

All experiments were carried out using Tesla V100S GPU with Pytorch~\cite{paszke2017automatic}. 
We use Adam optimizer with default $\beta_1, \beta_2$. The base learning rate is set to $0.02$ and halved at every 1000 iterations. Appropriate hyperparameters have been determined based on the performance of validation samples and are listed in Tables~\ref{t:hyperparameters}.
We also provide the codes and the trained models in the GitHub repository when the manuscript has been accepted.
\subsection{Figures of merit}
The performance is assessed both qualitatively and quantitatively. We use the peak-signal-to-noise ratio (PSNR) and the structural similarity index measure (SSIM). PSNR represents the ratio of the maximum possible value to the reconstruction error, which is defined as a log-scaled version of the mean squared error (MSE) between the reconstruction $\hat{\boldsymbol{x}}$ and the ground truth image $\boldsymbol{x}$,
$$
\operatorname{PSNR}\left(\hat{\boldsymbol{x}}, \boldsymbol{x} \right) =10 \log _{10}\left(\frac{L^{2}}{\operatorname{MSE}\left(\hat{\boldsymbol{x}}, \boldsymbol{x} \right)}\right),
$$
where the maximum image value $L=\max \left(\boldsymbol{x}\right)-\min \left(\boldsymbol{x}\right)$, and $\operatorname{MSE}\left(\hat{\boldsymbol{x}}, \boldsymbol{x}\right) =\frac{1}{n} \sum_{i=1}^{n}\left|\hat{x}_{i}-x_{i}\right|^{2}$. In general, a higher PSNR value indicates a better reconstruction quality. 

SSIM compares the overall structure of the image, including the luminance and contrast. It is based on assumptions about human visual perception. Results lie in the range $[0,1]$, with higher values being better. The SSIM is calculated through a sliding window at $M$ locations:
$$
\operatorname{\textbf{SSIM}}\left(\hat{\boldsymbol{x}}, \boldsymbol{x}\right) =\frac{1}{M} \sum_{j=1}^{M} \frac{\left(2 \hat{\mu}_{j} \mu_{j}+C_{1}\right)\left(2 \Sigma_{j}+C_{2}\right)}{\left(\hat{\mu}_{j}^{2}+\mu_{j}^{2}+C_{1}\right)\left(\hat{\sigma}_{j}^{2}+\sigma_{j}^{2}+C_{2}\right)}.
$$
Here, $\hat{\mu}_{j}$ and $\mu_{j}$ are the average pixel intensities, $\hat{\sigma}_{j}$ and $\sigma_{j}$ the variances and $\Sigma_{j}$ the covariance of $\hat{\boldsymbol{x}}$ and $\boldsymbol{x}$ at the $j$-th local window. Constants $C_{1}=\left(K_{1} L\right)^{2}$ and $C_{2}=\left(K_{2} L\right)^{2}$ stabilize the division. We use the same choices for the maximum image value $L$.

\subsection{Parallel beam Projection with Gaussian noise}
\subsubsection{Simulations}
We let $x$ denote the unknown image to be reconstructed. We assume $x$ is supported in the unit square $\Omega$ in $\mathbb{R}^{2}$ and we use a two-dimensional parallel beam CT. In this situation, the following Radon transform yields the projected data $f(s, \varphi)$ for each angle $\varphi \in[0, \pi]$ and distance from the origin $s \in \mathbb{R}$:
$$
f(s,\varphi) = Ax(s,\varphi) = \int_{\mathbb{R}} x\left(L_{s, \varphi}(t)\right) \mathrm{d} t,
$$
where $L_{s, \varphi}(t)$ is the path of an X-ray beam by the distance from $s$ and $\varphi$:
$$
L_{s, \varphi}(t) = s \boldsymbol{\theta} +t \boldsymbol{\theta} ^{\perp}, \qquad \boldsymbol{\theta}=(\cos \varphi, \sin \varphi).
$$
\zqp{In our experiments, we utilize a ground truth image of size $256 \times 256$. To generate the noisy measurement vector $y$, we add a vector $\epsilon \sim \mathcal{N}\left(0, \sigma^2\right)$ of Gaussian noise with a zero mean and a standard deviation of $\sigma$. The corresponding degradation model is as follows:
$$
y=f+\epsilon.
$$
We set $\sigma = 0.03$, which corresponds to a Signal-to-Noise Ratio (SNR) of $40 \mathrm{~dB}$.}
For the parallel Radon transform, we utilize 100 equispaced projections ranging from 0 to $\pi$.
In this setting, we employ the fidelity loss $\mathcal{F} (A \boldsymbol{x}, \boldsymbol{y})=\frac{1}{2}|A \boldsymbol{x}-\boldsymbol{y}|_{2}^{2}$.

\begin{figure}[!htb]
\centering
\includegraphics[width=0.8\textwidth]{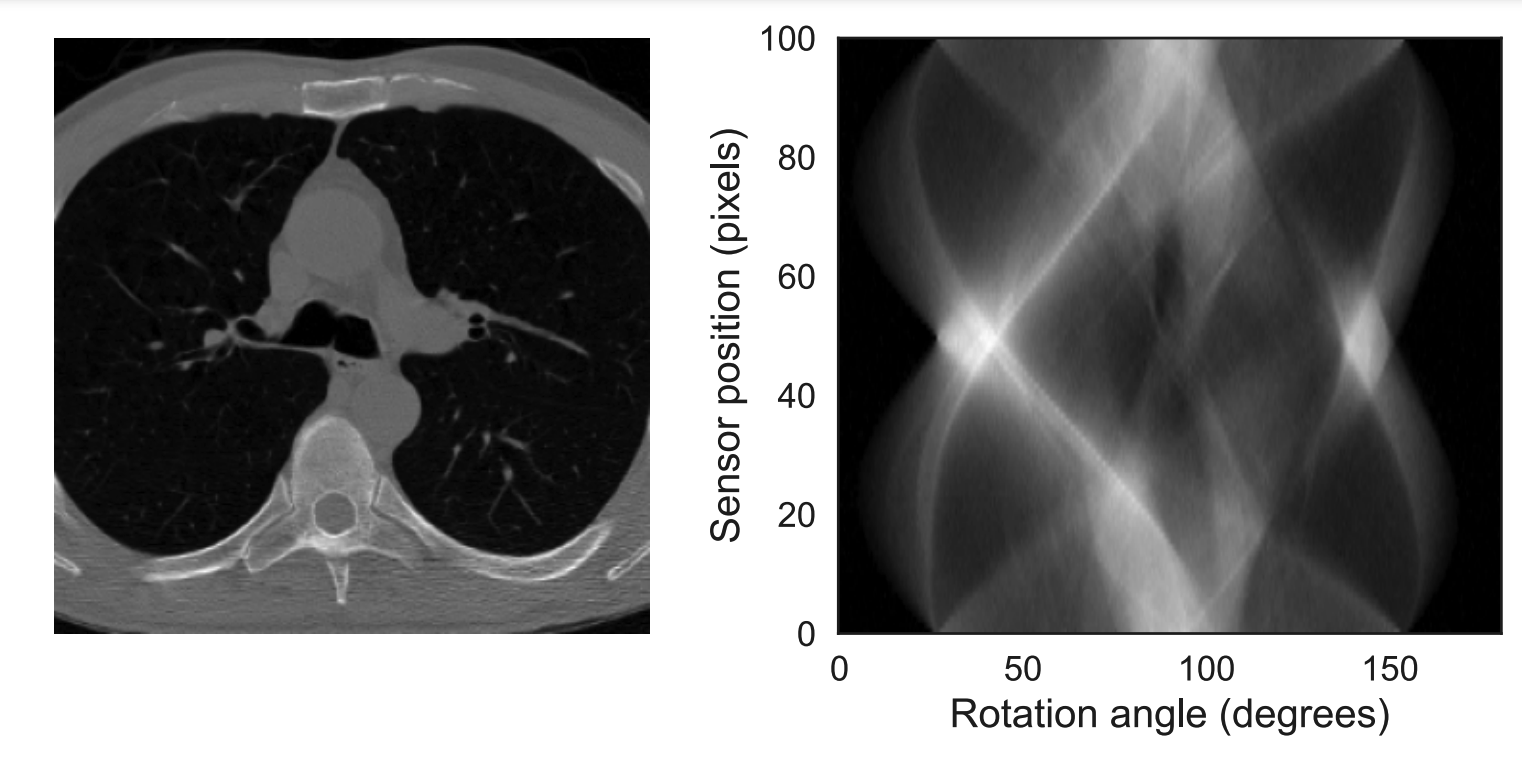}
\caption{((Gaussian noise) Left: the ground truth image. Right: the projection data simulated from the ground truth.}
\label{f:ct-truth-simu}
\end{figure}

\subsubsection{Benefit of the MCDIP scheme}
\zqp{To investigate the relationship between the number of codes and the performance of MCDIP-ADMM, we trained the model using an increasing number of latent codes for a test image, specifically $N=1, 5, 10, 20, 30$. It is important to note that when $N=1$, the MCDIP-ADMM model reverts to the PnP-DIP method. The regularizer parameters were exhaustively searched to maximize PSNR and exclude the impact of internal parameters. The results of our experiments are presented in Table~\ref{tab:latent_codes}. We observe a consistent improvement in PSNR as the number of codes increases, with the highest value achieved when the number of codes exceeds 20. However, it is worth noting that the PSNR values show relatively little change beyond a certain threshold, indicating that additional codes do not significantly enhance the quality of the reconstruction.
}
\begin{table}[htb!]
\centering
\caption{PSNR with different number of codes in MCDIP.}
\begin{tabular}{lllllll}
\hline
  $N$ & 1 & 5 & 10 & 20 & 30 \\ \hline
PSNR  & 26.15 & 28.27  & 30.21  &  30.51  & 30.52   \\ \hline
\end{tabular}
\label{tab:latent_codes}
\end{table}

\begin{figure}[!htb]
\centering
  \begin{subfigure}[t]{0.9\textwidth}
    \centering
    \includegraphics[width=0.98\linewidth]{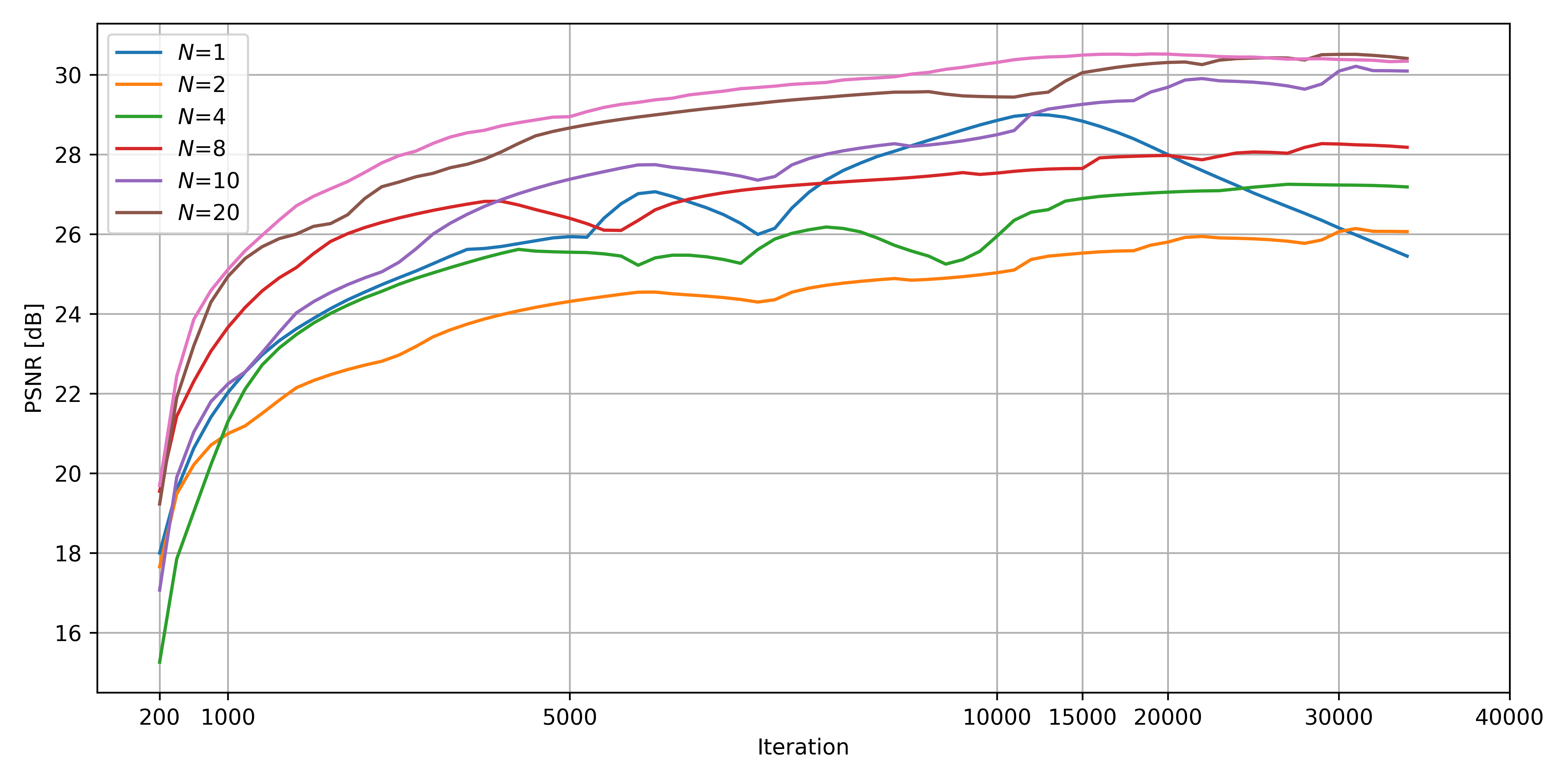}
  \end{subfigure}
\caption{Evolution of the PSNR with respect to the number of latent codes in MCDIP-ADMM. Each line represents the mean over 3 runs.}
\label{fig:psnr_2_iter}
\end{figure}

\zqp{Figure~\ref{fig:psnr_2_iter} illustrates the behavior of the peak signal-to-noise ratio (PSNR) observed in three separate runs. When $N=1$, the MCDIP-ADMM, reduced to the PnP-DIP method, exhibits a noticeable overfitting phenomenon, with the PSNR starting to decline after approximately 12,000 iterations. In contrast, our proposed multi-code approach successfully addresses this overfitting issue. The PSNR either maintains a consistent trend or shows improvement throughout the training iterations. This can be attributed to the enhanced ability of MCDIP to extract more comprehensive prior information, resulting in improved overall performance.}

\subsubsection{Performance comparison}
Table~\ref{t:ct_psnr_ssim_fl} presents the PSNR, SSIM, and fidelity loss of different methods. We compute the average PSNR improvement of three test images of the proposed method against its comparisons. 
The MCDIP-ADMM algorithm gives a 4.3 dB improvement over DIP, 1.7 dB over ADMM DIP-WTV, and 1.2 dB over PnP-DIP along with significantly higher SSIM values. 
It strongly shows that the MCDIP-ADMM model is superior to the other models in terms of PSNR and SSIM for all three test images. 
For image 2, the PSNR for the MCDIP-ADMM model is 32.45, compared to just 30.10 for PnP-DIP (the best one in comparison) which reconstructs images without multiple latent codes. 
It is worth noting that the gap between the final PSNR and the best PSNR of MCDIP-ADMM, which is 3.06, is significantly lower than that of PnP-DIP, which is 4.93. 
For images 1 and 3, similar conclusions can be obtained.
We can conclude that the proposed approach provides more accurate and robust restorations than its comparisons, largely thanks to its ability to extract multi-mode features from multiple latent codes.

Fidelity loss $\mathcal{F}\left(A \boldsymbol{x}, \boldsymbol{y} \right)$ in Eq.~\eqref{e:optim_orig} measures the consistency between the forward projected reconstruction and the observations. It is commonly used to check for data discrepancies and it can provide additional insights into the performance of the reconstruction methods. For an ideal model, the fidelity loss $\mathcal{F}\left(A \hat{\boldsymbol{x}}, \boldsymbol{y} \right)$ of the reconstruction $\hat{\boldsymbol{x}}$ would be close to the fidelity loss $\mathcal{F}\left(A {\boldsymbol{x}}, \boldsymbol{y} \right)$ of the truth $\boldsymbol{x}$. 
For image 1, ADMM DIP-WTV performs best in terms of fidelity loss, with a fidelity loss of 158.09, which is  extremely near to the fidelity loss of the true image. 
For image 2, the fidelity loss of ADMM DIP-WTV is slightly better than other approaches. 
For image 3, MCDIP-ADMM's fidelity loss is the closest to the truth fidelity loss, whereas the other techniques' fidelity losses are distant from truth fidelity loss.

\renewcommand{\arraystretch}{1.4}  
\begin{table}[!t]
\centering
\caption{Final (best) PSNR, SSIM and fidelity loss for different methods on three test images. The greater PSNR or SSIM, the better the reconstruction quality. And the better the reconstruction model, the closer the fidelity loss is to that of the true image. The best results are highlighted in bold. See Figs.~\ref{f:ct_gt2},~\ref{f:ct_gt0} and \ref{f:ct_gt1} for visualization.}
\resizebox{\textwidth}{!}{%
\begin{tabular}{rclrrcccc}
\hline
\multicolumn{1}{l}{}                                    & \begin{tabular}[c]{@{}c@{}}Image\end{tabular} & Truth  & \multicolumn{1}{l}{FBP} & \multicolumn{1}{l}{CGNE} & \multicolumn{1}{l}{DIP}                                   & \begin{tabular}[c]{@{}c@{}}ADMM \\ DIP-WTV\end{tabular}            & \multicolumn{1}{l}{PnP-DIP}                               & \multicolumn{1}{l}{MCDIP-ADMM}                                     \\ \hline
\multicolumn{1}{c}{}                                    & 1                                                     &        & 22.77                   & 25.09                    & \begin{tabular}[c]{@{}c@{}}26.14\\ (29.66)\end{tabular}   & \begin{tabular}[c]{@{}c@{}}29.34\\ (29.34)\end{tabular}            & \begin{tabular}[c]{@{}c@{}}29.91\\ (30.06)\end{tabular}   & \textbf{\begin{tabular}[c]{@{}c@{}}30.51\\ (30.51)\end{tabular}}   \\ \cline{2-9} 
PSNR                                                    & 2                                                     &        & 20.2                    & 22.56                    & \begin{tabular}[c]{@{}c@{}}27.97\\ (34.91)\end{tabular}   & \begin{tabular}[c]{@{}c@{}}30.06\\ (34.71)\end{tabular}            & \begin{tabular}[c]{@{}c@{}}30.1\\ (35.03)\end{tabular}    & \textbf{\begin{tabular}[c]{@{}c@{}}32.45\\ (35.51)\end{tabular}}   \\ \cline{2-9} 
                                                        & 3                                                     &        & 23.35                   & 25.87                    & \begin{tabular}[c]{@{}c@{}}27.82\\ (30.57)\end{tabular}   & \begin{tabular}[c]{@{}c@{}}30.30\\ (30.30)\end{tabular}            & \begin{tabular}[c]{@{}c@{}}31.26\\ (31.42)\end{tabular}   & \textbf{\begin{tabular}[c]{@{}c@{}}31.94\\ (31.94)\end{tabular}}   \\ \hline
                                                        & 1                                                     &        & 0.55                    & 0.66                     & \begin{tabular}[c]{@{}c@{}}0.74\\ (0.87)\end{tabular}     & \begin{tabular}[c]{@{}c@{}}0.87\\ (0.87)\end{tabular}              & \begin{tabular}[c]{@{}c@{}}0.87\\ (0.88)\end{tabular}     & \textbf{\begin{tabular}[c]{@{}c@{}}0.89\\ (0.89)\end{tabular}}     \\ \cline{2-9} 
SSIM                                                    & 2                                                     &        & 0.35                    & 0.47                     & \begin{tabular}[c]{@{}c@{}}0.77\\ (0.96)\end{tabular}     & \begin{tabular}[c]{@{}c@{}}0.84\\ (0.96)\end{tabular}              & \begin{tabular}[c]{@{}c@{}}0.85\\ (0.96)\end{tabular}     & \textbf{\begin{tabular}[c]{@{}c@{}}0.91\\ (0.96)\end{tabular}}     \\ \cline{2-9} 
                                                        & 3                                                     &        & 0.57                    & 0.69                     & \begin{tabular}[c]{@{}c@{}}0.82\\ (0.91)\end{tabular}     & \begin{tabular}[c]{@{}c@{}}0.91\\ (0.91)\end{tabular}              & \begin{tabular}[c]{@{}c@{}}0.91\\ (0.92)\end{tabular}     & \textbf{\begin{tabular}[c]{@{}c@{}}0.93\\ (0.93)\end{tabular}}     \\ \hline
                                                        & 1                                                     & 158.86 & 118.69                  & 86.63                    & \begin{tabular}[c]{@{}c@{}}134.74\\ (134.74)\end{tabular} & \textbf{\begin{tabular}[c]{@{}c@{}}158.09\\ (158.09)\end{tabular}} & \begin{tabular}[c]{@{}c@{}}144.24\\ (144.16)\end{tabular} & \begin{tabular}[c]{@{}c@{}}151.26\\ (151.26)\end{tabular}          \\ \cline{2-9} 
\begin{tabular}[c]{@{}r@{}}Fidelity \\ Loss\end{tabular} & 2                                                     & 203.25 & 148.54                  & 112.46                   & \begin{tabular}[c]{@{}c@{}}181.02\\ (170.64)\end{tabular} & \textbf{\begin{tabular}[c]{@{}c@{}}183.54\\ (183.54)\end{tabular}} & \begin{tabular}[c]{@{}c@{}}174.33\\ (174.26)\end{tabular} & \begin{tabular}[c]{@{}c@{}}179.10\\ (175.57)\end{tabular}          \\ \cline{2-9} 
\multicolumn{1}{l}{}                                    & 3                                                     & 156.19 & 117.87                  & 85.83                    & \begin{tabular}[c]{@{}c@{}}135.49\\ (135.49)\end{tabular} & \begin{tabular}[c]{@{}c@{}}145.13\\ (145.13)\end{tabular}          & \begin{tabular}[c]{@{}c@{}}141.50\\ (141.50)\end{tabular} & \textbf{\begin{tabular}[c]{@{}c@{}}150.05\\ (150.05)\end{tabular}} \\ \hline
\end{tabular}%
}
\label{t:ct_psnr_ssim_fl}
\end{table}

\begin{figure}[!htb]
\centering
\includegraphics[width=0.85\textwidth]{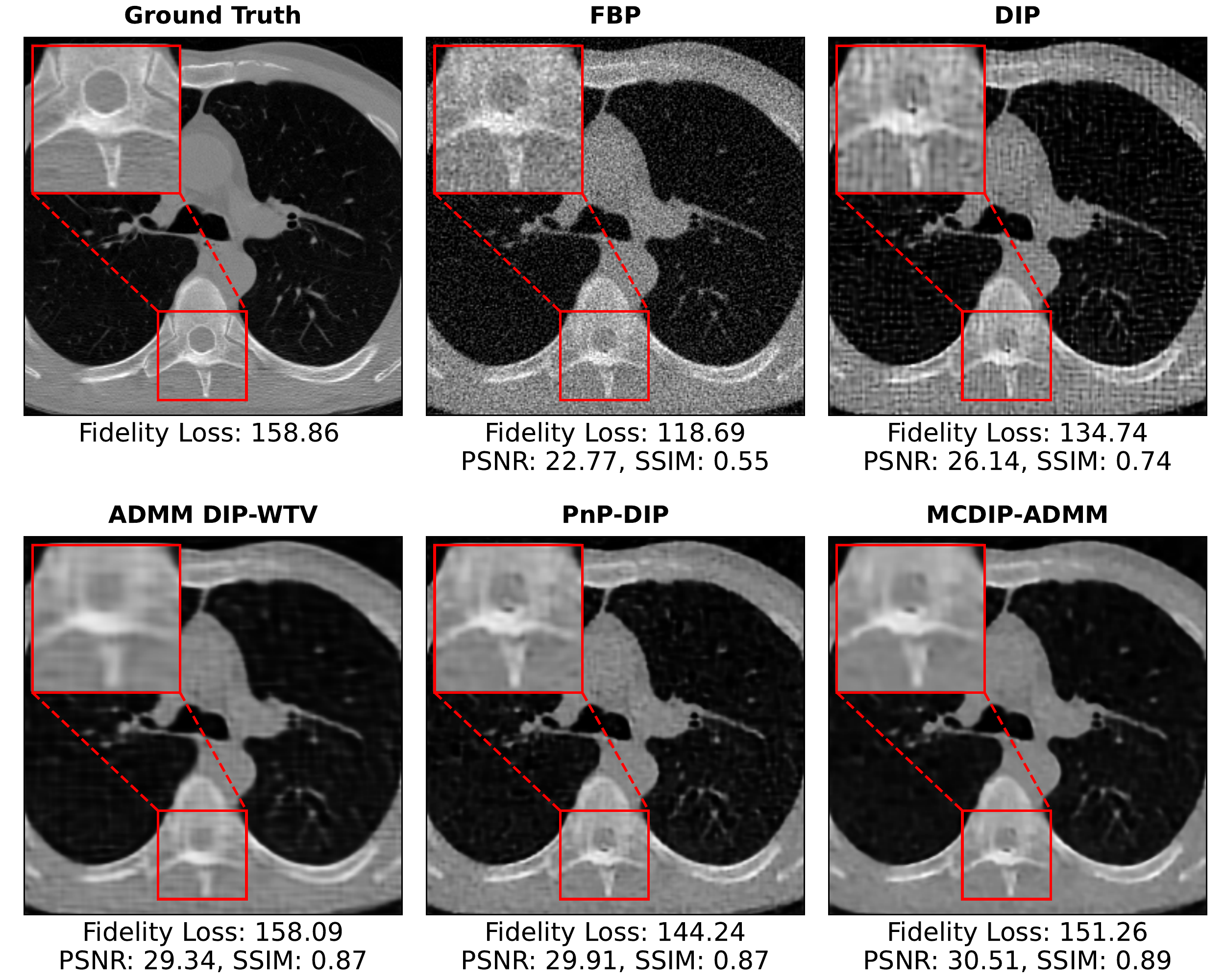}
\caption{Reconstruction of image 1: Comparison of reconstructions obtained using FBP, DIP, \tclr{ADMM DIP-WTV}, PnP-DIP, and MCDIP-ADMM. The fidelity loss, PSNR, and SSIM values are displayed below the picture, where fidelity loss measures the discrepancy between the observations and the noise-free projection of the reconstructed image.}
\label{f:ct_gt2}
\end{figure}

\begin{figure}[!htb]
\centering
\includegraphics[width=0.85\textwidth]{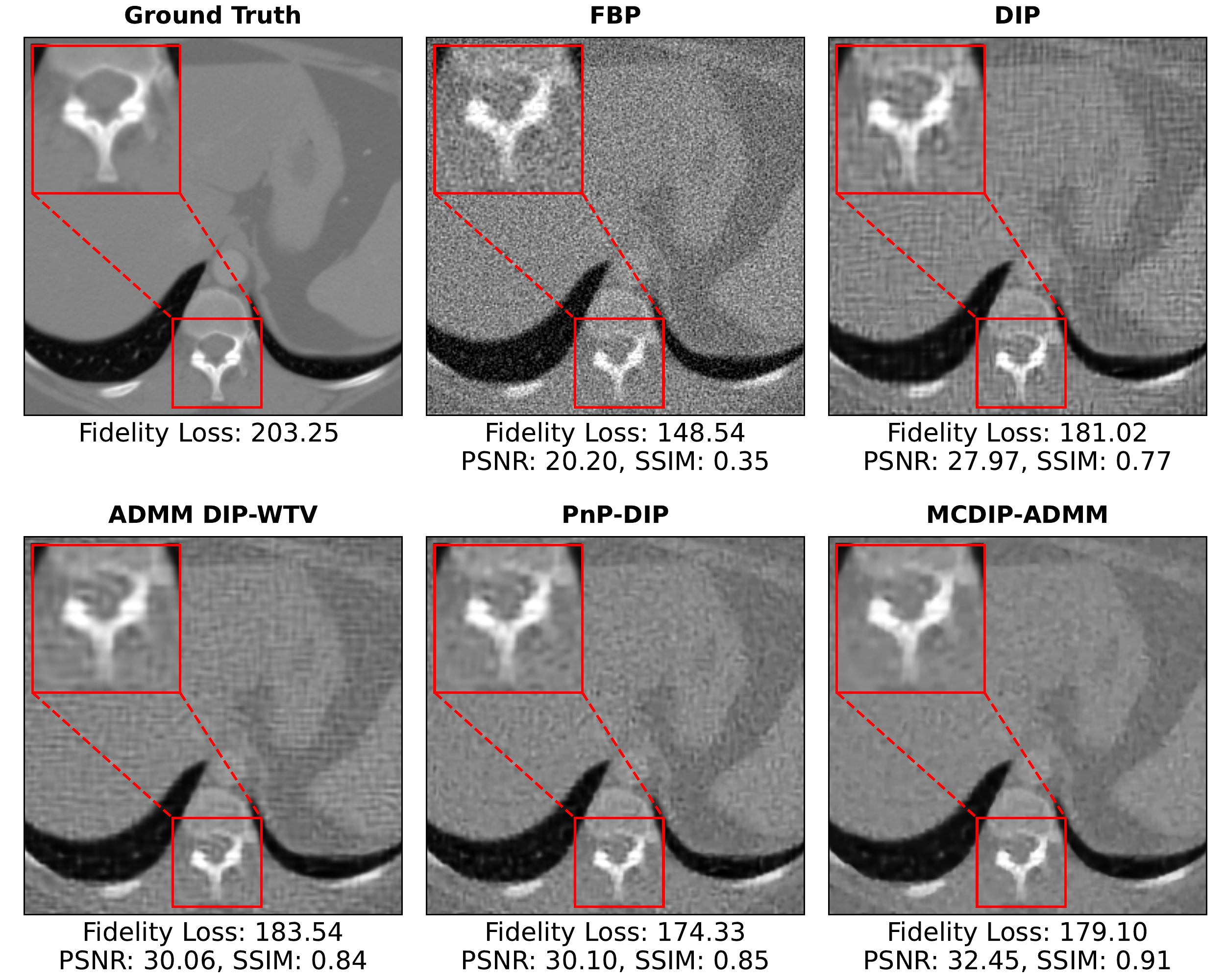}
\caption{Reconstruction of image 2: Comparison of reconstructions obtained using FBP, DIP, ADMM DIP-WTV, PnP-DIP, and MCDIP-ADMM. The fidelity loss, PSNR, and SSIM values are displayed below the picture, where fidelity loss measures the discrepancy between the observations and the noise-free projection of the reconstructed image.}
\label{f:ct_gt0}
\end{figure}

\begin{figure}[!htb]
\centering
\includegraphics[width=0.85\textwidth]{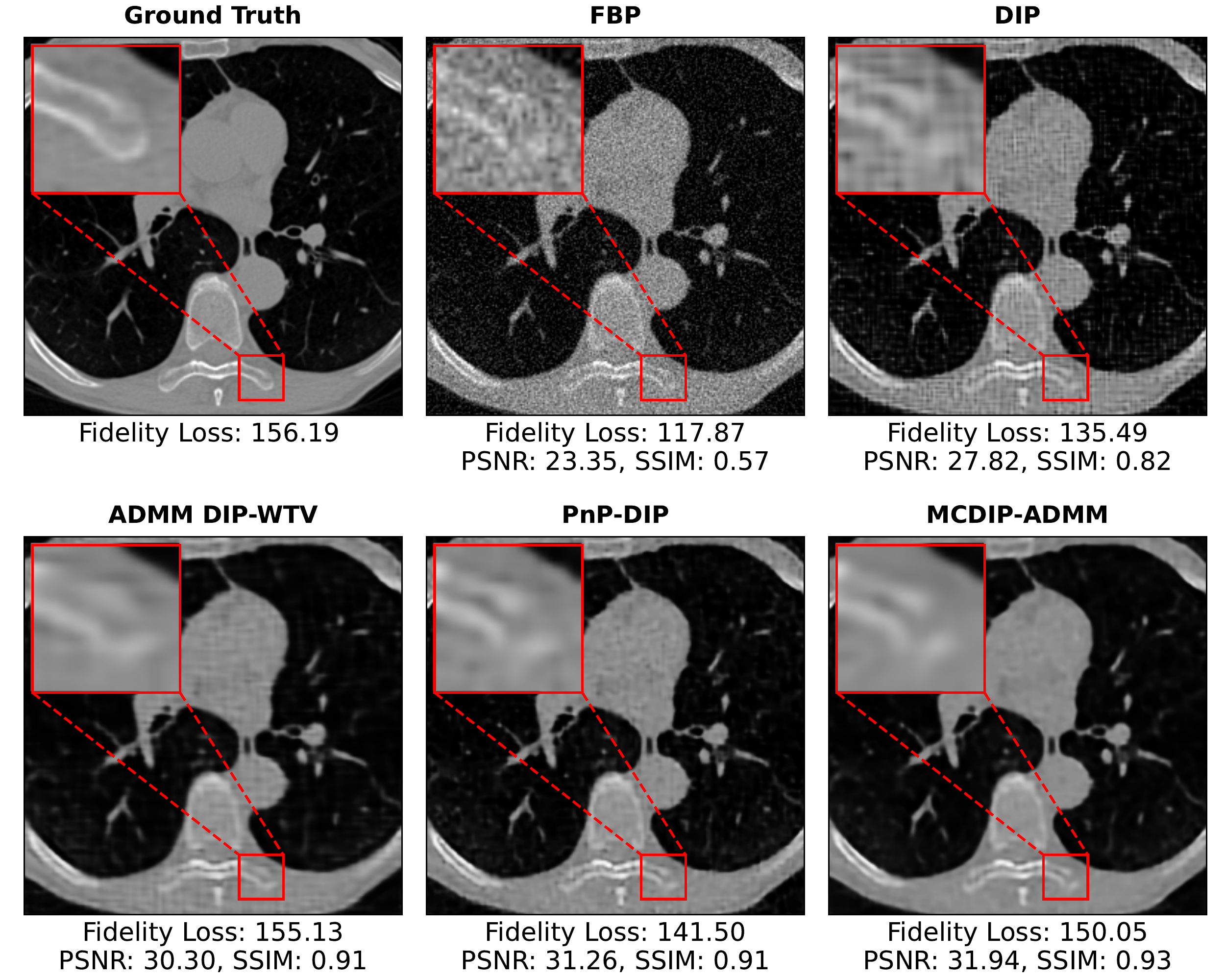}
\caption{Reconstruction of image 3: Comparison of reconstructions obtained using FBP, DIP, ADMM DIP-WTV, PnP-DIP and MCDIP-ADMM. The fidelity loss, PSNR, and SSIM values are displayed below the picture, where fidelity loss measures the discrepancy between the observations and the noise-free projection of the reconstructed image.}
\label{f:ct_gt1}
\end{figure}

\renewcommand{\arraystretch}{1}
Figs.~\ref{f:ct_gt2}, \ref{f:ct_gt0}, and \ref{f:ct_gt1} provide a representative reconstruction  for three test images using FBP, DIP, ADMM DIP-WTV, PnP-DIP, and the proposed method. 
Our reconstruction is of the greatest quality, both visually and in terms of SSIM and PSNR. We find overall noise pollution in the images obtained using the FBP and {vanilla DIP} approaches. Many artifacts can be seen in the DIP restoration as a result of {severe overfitting}. The fidelity loss of {the ADMM DIP-WTV} and MCDIP-ADMM is close to that of the ground truth, {indicating that both methods mitigate overfitting}. Other methods, including PnP-DIP, have terrible data consistency since their fidelity loss is significantly smaller than the ground truth.
Furthermore, because the prior's multi-encoding structure can collect more latent information of an image, this method is faster to achieve the maximum plateau region. 

\clearpage
\subsection{Fan-beam projection with Poisson noise}
\subsubsection{Simulations}
\begin{figure}[!htb]
\centering
\includegraphics[width=0.9\textwidth]{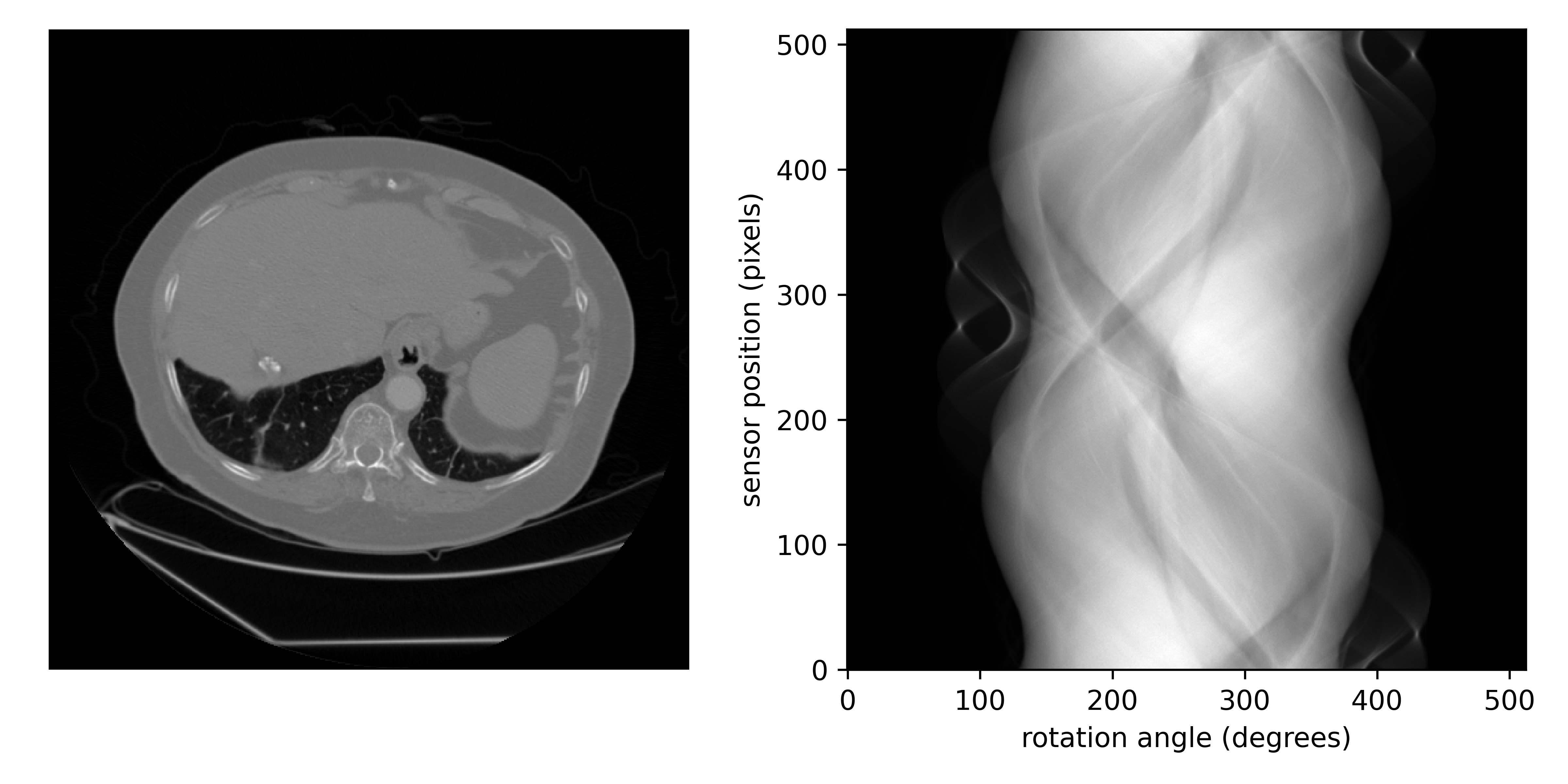}
\caption{(Poisson noise) Left: the ground truth image. Right: the projection data simulated from the ground truth.}
\label{f:truth_sino_2_poisson}
\end{figure}

\zqp{
We utilize a ground truth image of size $512 \times 512$. We simulate fan-beam projections with Poisson noise, i.e., the full model is given by
\begin{equation}\label{eq:poisson_forward}
    \pi({y}| x)=\pi_{\mathrm{P}}({y}| {f}=\mathrm{A} x),
\end{equation}
where $\pi_{\mathrm{P}}({y}|{f})$ represents the $d$-dimensional Poisson distribution:
\begin{equation}\label{eq:ct_poisson}
     \pi_{\mathrm{P}}({y} | {f})=\prod_{i=1}^{m} \frac{\left(f_{i}\right)^{y_{i}} \exp \left(-f_{i}\right)}{y_{i} !}.
\end{equation}
In our numerical experiments, we employ 512 sampling angles uniformly distributed in the interval $[0, 2\pi)$, position the X-ray source at a distance of 512 units, and place the detector at the same distance of 512 units.
The detector consists of 512 pixels with a spacing of 2.0 units, ensuring that the rays cover the entire image~\cite{ronchetti2020torchradon}.
Figure~\ref{f:truth_sino_2_poisson} presents the ground truth image and its corresponding sinogram. To ensure consistency with the Poisson noise model, we employ an appropriate Poisson regression loss as our data fidelity loss, which maximizes the likelihood under this model. The loss function is defined as follows:
$$\mathcal{F} (A {x}, {y})=\langle\mathbf{1}, A {x}\rangle-\langle{y}, \log (A {x})\rangle.$$ 
}

\subsubsection{Performance comparison}
\begin{figure}[!htb]
\centering
\includegraphics[width=0.75\textwidth]{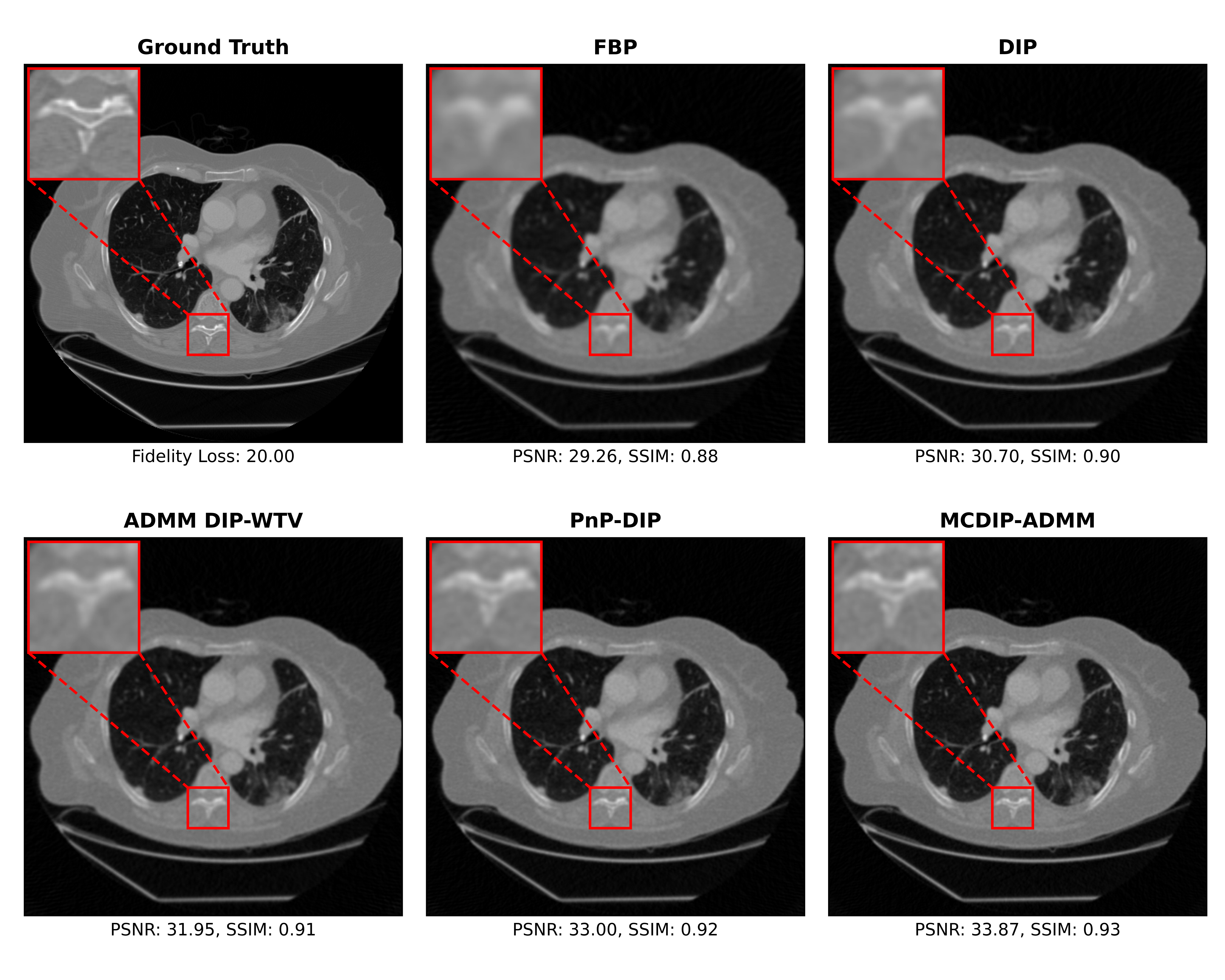}
\caption{Reconstruction of image 1: Comparison of reconstructions obtained using FBP, DIP, ADMM DIP-WTV, PnP-DIP, and MCDIP-ADMM. The PSNR and SSIM values are displayed below the picture.}
\label{f:ct_gt2_poisson}
\end{figure}

\begin{figure}[!htb]
\centering
\includegraphics[width=0.75\textwidth]{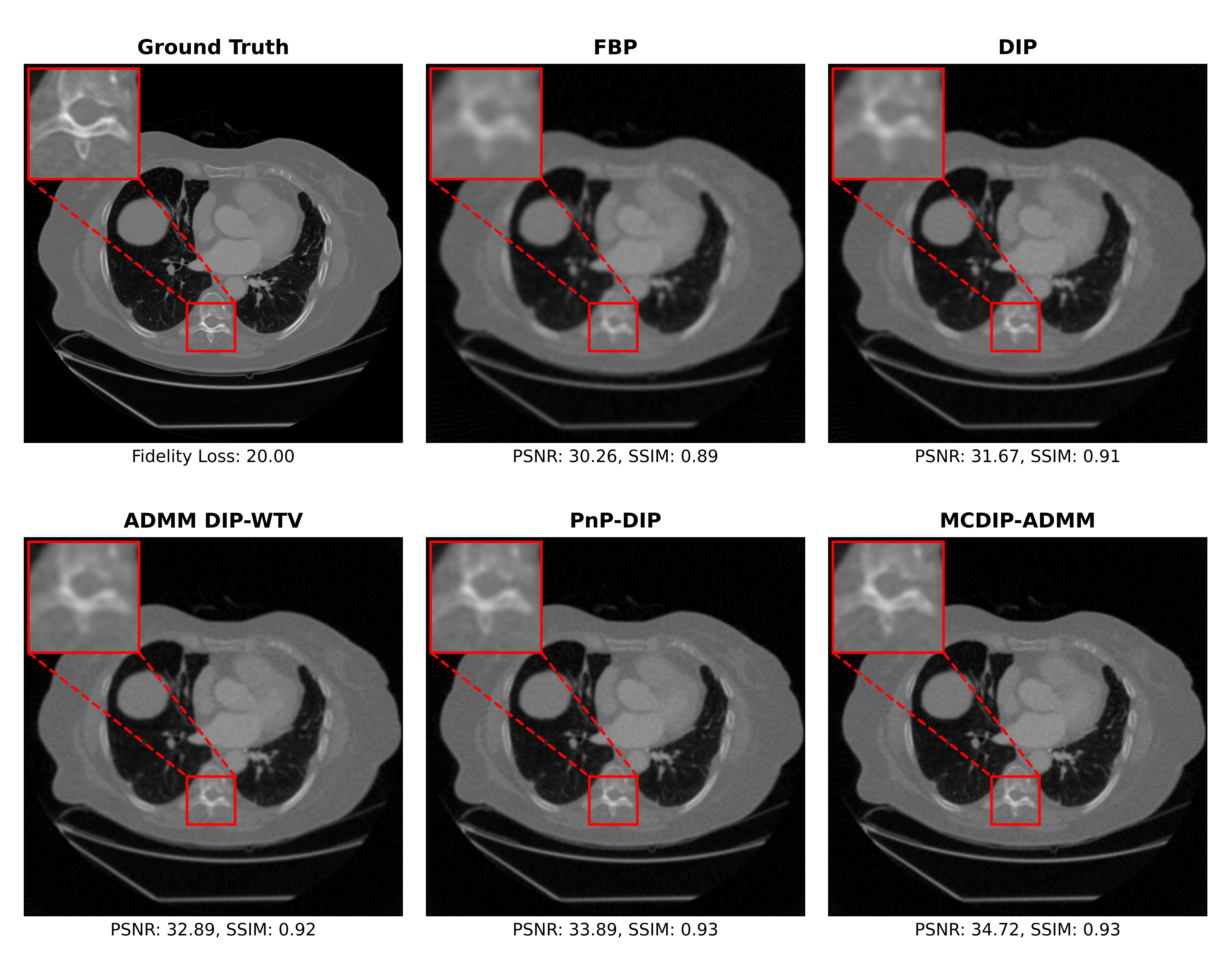}
\caption{Reconstruction of image 2: Comparison of reconstructions obtained using FBP, DIP, ADMM DIP-WTV, PnP-DIP, and MCDIP-ADMM. The PSNR and SSIM values are displayed below the picture.}
\label{f:ct_gt0_poisson}
\end{figure}

\begin{figure}[!htb]s
\centering
\includegraphics[width=0.75\textwidth]{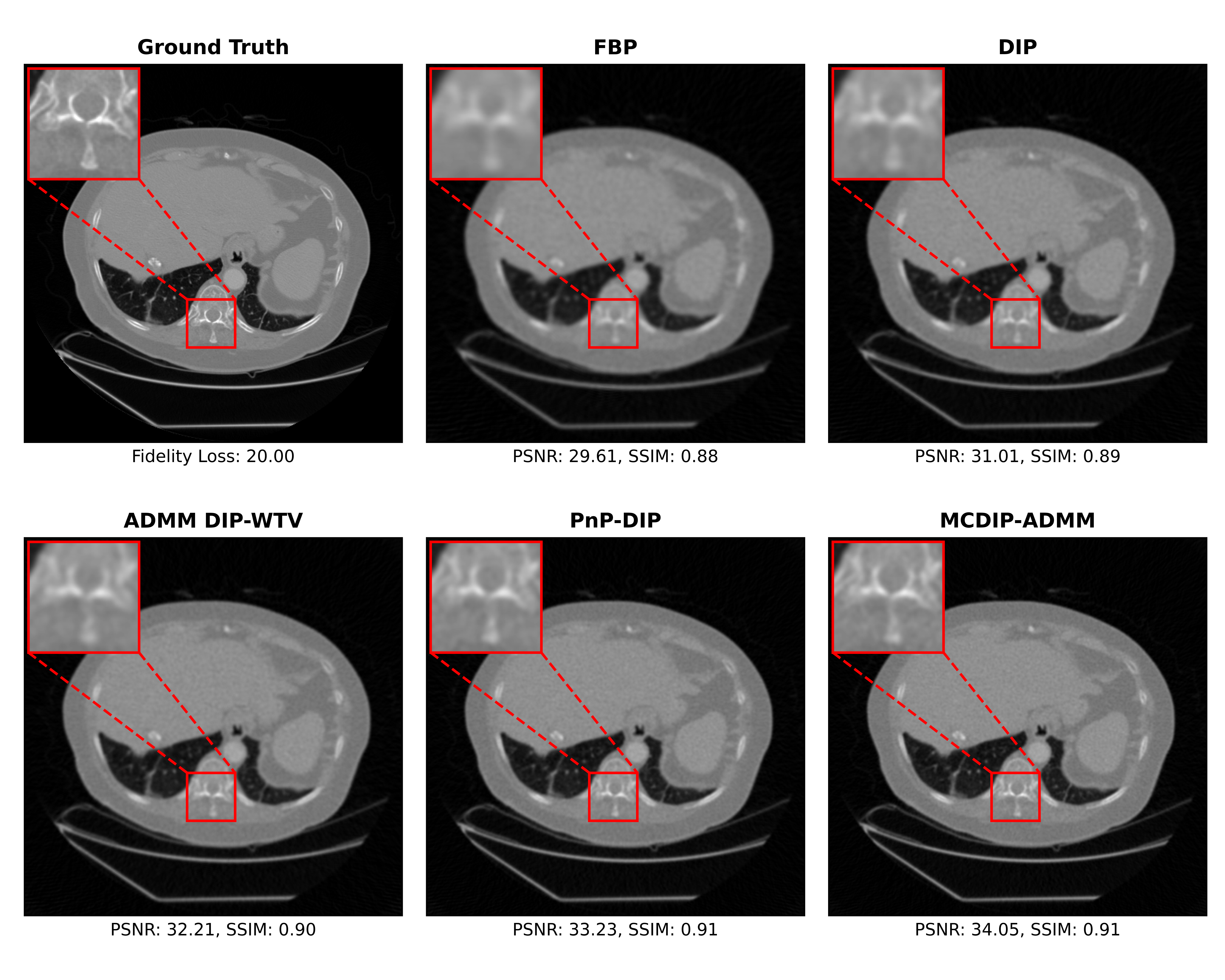}
\caption{Reconstruction of image 3: Comparison of reconstructions obtained using FBP, DIP, ADMM DIP-WTV, PnP-DIP, and MCDIP-ADMM. The PSNR and SSIM values are displayed below the picture.}
\label{f:ct_gt1_poisson}
\end{figure}

\zqp{
Table~\ref{t:ct_psnr_ssim_poisson} summarizes the numerical results of all competing methods. 
We conjecture that similar outcomes can be achieved in parallel beam projection settings as well. The MCDIP scheme demonstrates its ability to automatically extract multi-mode features from multiple latent codes, thereby offering a more efficient and effective approach.
Our observations reveal that the MCDIP-ADMM algorithm outperforms other methods, with a notable improvement of 3.09 dB over DIP, 1.86 dB over ADMM DIP-WTV, and 0.84 dB over PnP-DIP, while also yielding significantly higher SSIM values. Visual results of Figures~\ref{f:ct_gt2_poisson}, \ref{f:ct_gt0_poisson}, and \ref{f:ct_gt1_poisson} confirm that our reconstruction achieves the highest quality both visually and in terms of SSIM and PSNR.
}

\renewcommand{\arraystretch}{1.4}  
\begin{table}[!htb]
\centering
\caption{Final (best) PSNR, SSIM, and fidelity loss for different methods on three test images. The greater PSNR or SSIM, the better the reconstruction quality. Noting that we do not present fidelity loss due to their substantial variation under Poisson noise conditions. The best results are highlighted in bold. See Figs.~\ref{f:ct_gt2_poisson},~\ref{f:ct_gt0_poisson} and \ref{f:ct_gt1_poisson} for visualization.}
\resizebox{\textwidth}{!}{%
\begin{tabular}{rcrcccc}
\hline
\multicolumn{1}{l}{} & Image & \multicolumn{1}{l}{FBP} & \multicolumn{1}{l}{DIP}                                                   & ADMM DIP-WTV                                             & \multicolumn{1}{l}{PnP-DIP}                                    & \multicolumn{1}{l}{MCDIP-ADMM}                                   \\ \hline
\multicolumn{1}{c}{} & 1     & 29.61                   & \begin{tabular}[c]{@{}c@{}}31.01\\ (32.56)\end{tabular}                   & \begin{tabular}[c]{@{}c@{}}32.21\\ (32.76)\end{tabular}  & \begin{tabular}[c]{@{}c@{}}33.23\\ (33.92)\end{tabular}        & \textbf{\begin{tabular}[c]{@{}c@{}}34.05\\ (34.08)\end{tabular}} \\ \cline{2-7} 
PSNR                 & 2     & 29.26                   & \begin{tabular}[c]{@{}c@{}}30.70\\ (32.02)\end{tabular}                   & \begin{tabular}[c]{@{}c@{}}31.95\\ (32.73)\end{tabular}  & \begin{tabular}[c]{@{}c@{}}33.00\\ (33.87)\end{tabular}        & \textbf{\begin{tabular}[c]{@{}c@{}}33.87\\ (33.92)\end{tabular}} \\ \cline{2-7} 
                     & 3     & 30.26                   & \begin{tabular}[c]{@{}c@{}}31.67\\ (33.04)\end{tabular}                   & \begin{tabular}[c]{@{}c@{}}32.89\\ (33.93))\end{tabular} & \begin{tabular}[c]{@{}c@{}}33.89\\ (34.23)\end{tabular}        & \textbf{\begin{tabular}[c]{@{}c@{}}34.72\\ (34.82)\end{tabular}} \\ \hline
                     & 1     & 0.88                    & \begin{tabular}[c]{@{}c@{}}0.89\\ (0.91)\end{tabular}                     & \begin{tabular}[c]{@{}c@{}}0.90\\ (0.91)\end{tabular}    & \textbf{\begin{tabular}[c]{@{}c@{}}0.91\\ (0.93)\end{tabular}} & \textbf{\begin{tabular}[c]{@{}c@{}}0.91\\ (0.94)\end{tabular}}   \\ \cline{2-7} 
SSIM                 & 2     & 0.88                    & \multicolumn{1}{r}{\begin{tabular}[c]{@{}r@{}}0.90\\ (0.92)\end{tabular}} & \begin{tabular}[c]{@{}c@{}}0.91\\ (0.93)\end{tabular}    & \textbf{\begin{tabular}[c]{@{}c@{}}0.92\\ (0.95)\end{tabular}} & \textbf{\begin{tabular}[c]{@{}c@{}}0.92\\ (0.95)\end{tabular}}   \\ \cline{2-7} 
                     & 3     & 0.89                    & \multicolumn{1}{r}{\begin{tabular}[c]{@{}r@{}}0.91\\ (0.94)\end{tabular}} & \begin{tabular}[c]{@{}c@{}}0.92\\ (0.93)\end{tabular}    & \textbf{\begin{tabular}[c]{@{}c@{}}0.93\\ (0.94)\end{tabular}} & \textbf{\begin{tabular}[c]{@{}c@{}}0.93\\ (0.93)\end{tabular}}   \\ \hline
\end{tabular}
}
\label{t:ct_psnr_ssim_poisson}
\end{table}

\clearpage
\section{Conclusion}\label{sec:conclusion} 
\zqp{In this study, we propose an unsupervised method called MCDIP-ADMM, aimed at addressing the issue of overfitting in DIP-based CT inverse problems. Our method incorporates a multi-Code Deep Image Prior (MCDIP) and utilizes the plug-and-play alternative direction method of multipliers (ADMM). Through extensive testing, we demonstrate that our approach achieves superior accuracy compared to three other competing methods, namely DIP, PnP-DIP, and ADMM DIP-WTV. This conclusion is supported by experiments conducted under two different configurations: parallel beam projection with Gaussian noise and fan-beam projection with Poisson noise.}

A main perspective for future work is to extend the proposed method to a Bayesian framework, for example, by developing a Bayesian approach~\cite{cheng2019bayesian,kohl2018probabilistic}, and by replacing the standard DIP with a generative model where the prior has an explicit mathematical form, as in~\cite{denker2021conditional}. 
That work takes conditional normalizing flow to reconstruct the image, but it would be interesting to combine it with optimization algorithms.  

\clearpage
\section*{Appendix}
\setcounter{table}{0}
\renewcommand{\thetable}{A\arabic{table}}
\parindent 0pt
\renewcommand{\arraystretch}{1.4}  
We present all the hyperparameters in Table~\ref{t:hyperparameters}. 
\begin{table}[!htb]
\centering
\caption{Hyperparameters of CT image reconstruction with parallel beam projection (Gaussian noise) and fan-beam projection (Poisson noise).}
\label{t:hyperparameters}
\begin{tabular}{c|c|cc}
\hline
\multicolumn{1}{l|}{}                                                         & \multicolumn{1}{l|}{} & \multicolumn{1}{c|}{Fan-beam projection }                                                                  & Parallel beam projection                                                                  \\ \hline
\multirow{2}{*}{\begin{tabular}[c]{@{}c@{}}PnP-DIP\\ MCDIP-ADMM\end{tabular}} & $\rho$                & \multicolumn{2}{c}{1}                                                                                                                                                                \\ \cline{2-4} 
                                                                              & $\lambda$             & \multicolumn{1}{c|}{8.0}                                                                       & 4.0                                                                                \\ \hline
\multicolumn{1}{l|}{MCDIP-ADMM}                                               & $N$                   & \multicolumn{1}{c|}{15}                                                                         & 20                                                                                 \\ \hline
\multirow{2}{*}{DIP-VBTV}                                                     & $\lambda$             & \multicolumn{2}{c}{0.05}                                                                                                                                                             \\ \cline{2-4} 
                                                                              & other                 & \multicolumn{2}{l}{\begin{tabular}[c]{@{}l@{}}Color sapce: Eq.(2.24) in~\cite{batard2021dip}\\      h: diag(900,3000,3000)\\       g: Eq.(2.9) in~\cite{batard2021dip}\end{tabular}} \\ \hline
\end{tabular}
\end{table}

\section*{Acknowledgements} 
The work is supported by the National Natural Science Foundation of China under Grant 12101614 and the Natural Science Foundation of Hunan Province, China, under Grant 2021JJ40715. We are grateful to the High Performance Computing Center of Central South University for assistance with the computations.




\end{document}